\documentclass{jaa}
\usepackage{natbib}
\usepackage{graphicx}

\begin{document}

\title{Constraining the onset height of coronal mass ejection driven shocks 
using near-Sun observations in visible and radio wavelengths}


\author{C. Kathiravan, V. Muthupriyal, and R. Ramesh}
\affilOne{Indian Institute of Astrophysics, Bangalore 560 034, Karnataka, India.\\}



\twocolumn[{

\maketitle

\corres{kathir@iiap.res.in}


\begin{abstract}
One of the debated issues about the onset of the type II radio bursts near the Sun is the heliocentric distance ($r$) at which the associated magnetohydrodynamic (MHD) shocks are formed, and the association of the latter with the coronal mass ejections (CMEs). The debate is primarily due to the absence of routine CME observations in whitelight at 
$r\,{<}\,1.5R_{\odot}$. We present here an example for how joint observations with the  Visible Emission Line Coronagraph (VELC) onboard the recently launched ADITYA-L1 (the first dedicated Indian space solar mission, \citealp{Parate2025}), and Gauribidanur radio facilities could be useful to address the issue.
\end{abstract}


}]


\pgrange{1--}
\setcounter{page}{1}
\lp{8}

\section{Introduction}
Transients like the coronal mass ejections (CME) generate and drive magntohydrodynamic (MHD) shocks as they propagate outwards in the solar atmosphere. The shocks accelerate electrons. The plasma oscillations excited by the accelerated electrons generate escaping radio waves. The spectral observations of type II radio bursts with relatively slow drift (${\approx}$0.5\,MHz\,$\rm s^{-1}$) from the high to low frequencies in the time-frequency plane are the signatures of such shocks \citep{Nelson1985,Gopalswamy2006}. The bursts are due to plasma emission. At times, both the fundamental (F) and harmonic (H)  components of the bursts, at the local electron plasma frequency of the propagating shock, can be seen as two emission bands with a
frequency ratio of ${\approx}$1:2 \citep{Wild1954,Zlotnik1998}. The drift of the bursts from high to low frequencies with time is due to the decrease of electron density ($N_{e}$) with increasing $r$ in the solar atmosphere, and radio emission from successive plasma levels of decreasing $N_{e}$ due to the outward propagating CME/shock. The radio emission can be observed at various distances from the Sun, starting from the inner corona to the orbit of Earth and beyond, depending on the shock propagation. 
CME shocks are closely associated with Ground Level Enhancement (GLE) in Solar Energetic Particle (SEP events), as well as Storm Sudden Commencement (SSC) events at Earth also \citep{Gopalswamy2012a,Veenadhari2012}. The CME height at the onset time of a type II burst indicates the distance at which the CME becomes super‐Alfv{\'e}nic to drive a MHD shock. 

The start frequency of coronal type II radio bursts is mostly in the range 200\,-\,50 MHz. The plasma levels corresponding to the above frequency range are typically located at 
$r{\approx}$1.2\,-\,2.0$R_{\odot}$ \citep{Gopalswamy2005}. Routine observations of CMEs, particularly with the Large Angle and Spectrometric Coronagraphs C2 \& C3 (LASCO C2 \& C3) onboard the Solar and Heliospheric Observatory (SoHO) are in the range $r{\approx}$2.2\,-\,30$R_{\odot}$. The lower limit is due to the size of the coronagraph occulter \citep{Brueckner1995}. Due to this larger size, the location of the leading edge of the CMEs at the time of onset of the coronal type II bursts are mostly inferred from the extrapolated height-time plot of the CMEs. With the VELC \citep{Prasad2017,Prasad2023,Singh2019,Singh2025}, it has now become possible to infer the time at which the CMEs were observed very close to the Sun 
($r{\approx}1.05R_{\odot}$). We present one such near-Sun observation of a CME around the same time when a type II solar radio burst was observed with the facilities in the Gauribidanur observatory \citep{Ramesh2011a,Ramesh2014}, and show that CMEs can drive MHD shocks as close as $r{\approx}1.19R_{\odot}$ in the solar atmosphere. 

\section{Observations}

\begin{figure*}
\centerline{\includegraphics[height=0.35\textheight]{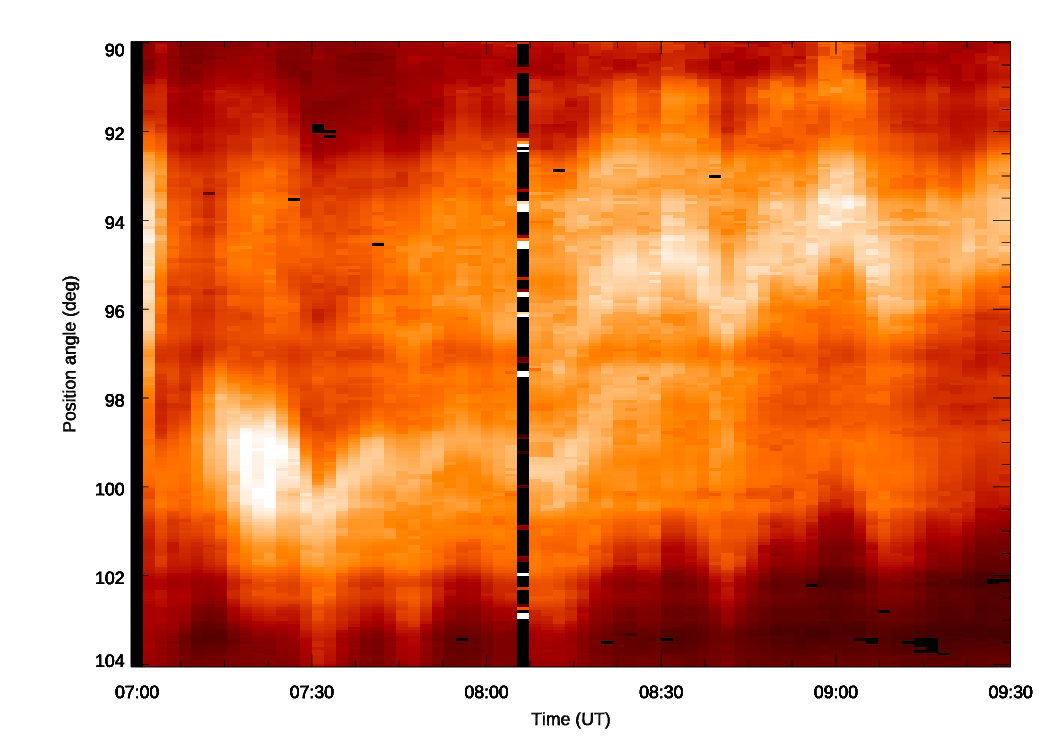}}
\caption{Brightness of the 5303{\AA} emission line from the solar corona observed along the 1st slit in VELC/ADITYA-L1 as a function of time on 2024 May 27. The position angle range of observations shown is ${\approx}\,90^{\circ}$\,-\,$104^{\circ}$. The enhanced emission near position angle 
${\approx}\,100^{\circ}$ whose onset is at ${\approx}$07:04\,UT is discussed in the main text. The other enhancement near position angle 
${\approx}\,95^{\circ}$ with onset at ${\approx}$08:15\,UT is due to coronal loop activity. The vertical black patch near 08:00\,UT is due to data error. The VELC data is available from 07:01\,UT only. Hence there is a gap in the beginning.} 
\label{figone}
\end{figure*}

\subsection{VELC/ADITYA-L1}
The VELC payload onboard ADITYA-L1 is an internally occulted solar coronagraph with capability to carry out simultaneous imaging (5000{\AA}) and spectroscopy (5303{\AA}, 7892{\AA}, \& 10747{\AA}) observations close to the limb. 
The radius of the occulter is $1.05R_{\odot}$.  The field-of-view (FoV) for the imaging and spectral channels are 1.05\,-\,3.0$R_{\odot}$ and 
1.05\,-\,1.5$R_{\odot}$, respectively. The spectral observations are carried out using four straight slits simultaneously. The use of multi-slits helps to 
reduce the observing time in the spectral channels by a factor of four, particularly when entire FoV is observed. The length of each slit is along the north-south direction of Sun, and width (dispersion) is along the east-west direction of Sun. The FoV covered by the slit length is ${\pm}$1.5$R_{\odot}$.
Typically, the 1st and 4th slits are used to observe the corona above the east and west limbs of the Sun, respectively. The coronal regions above the north and south limbs of the Sun are observed using the 2nd and 3rd slits together. 

The observations can be carried out in either sit \& stare mode or raster scan mode \citep{Prasad2023,Singh2025,Ramesh2024,Muthupriyal2025}.
The slit width (dispersion direction) is 50${\mu}$m, equivalent to 
$9.6^{\prime\prime}$.
A linear scan mechanism (LSM) is used to position the coronal light on the slits.  
The 2nd and 3rd slits are equi-distant in between the 1st and 4th slits. The spacing between the adjacent slits is
$0.75R_{\odot}$. A distance range of $0.75R_{\odot}$ in the east-west direction of the Sun can be scanned (observed) using each slit,
by moving the LSM. The total FoV that can be observed in the east-west direction using  the four slits together is ${\pm}$1.5$R_{\odot}$. 
The observations reported were carried out with the 5303\,{\AA} (green line) spectroscopy channel on 2024 May 27 during 07:00\,UT\,-\,09:30\,UT in the sit and stare mode.
Considering the occulter size in the VELC and the size of Sun's image as observed from the Sun-Earth Lagrangian L1 location on 2024 May 27,
we find that the occulter edge will 
be at $r\,{\approx}\,1.13R_{\odot}$ in the corona.
The LSM was positioned such that the center of the 1st slit observes the corona at 
$r$\,=\,1.13$R_{\odot}$ to the east of the Sun's central meridian. The spectra were obtained with an exposure time of 5\,s at a cadence of 123\,s. 
Each spectrum was corrected for dark current, curvature of the spectra, flat-field, and background as described in \cite{Singh2025}. Then 
the observed emission line profile was fitted with a Gaussian to compute the peak intensity at each spatial location along the slit.

\begin{figure}
\includegraphics[width=\columnwidth]{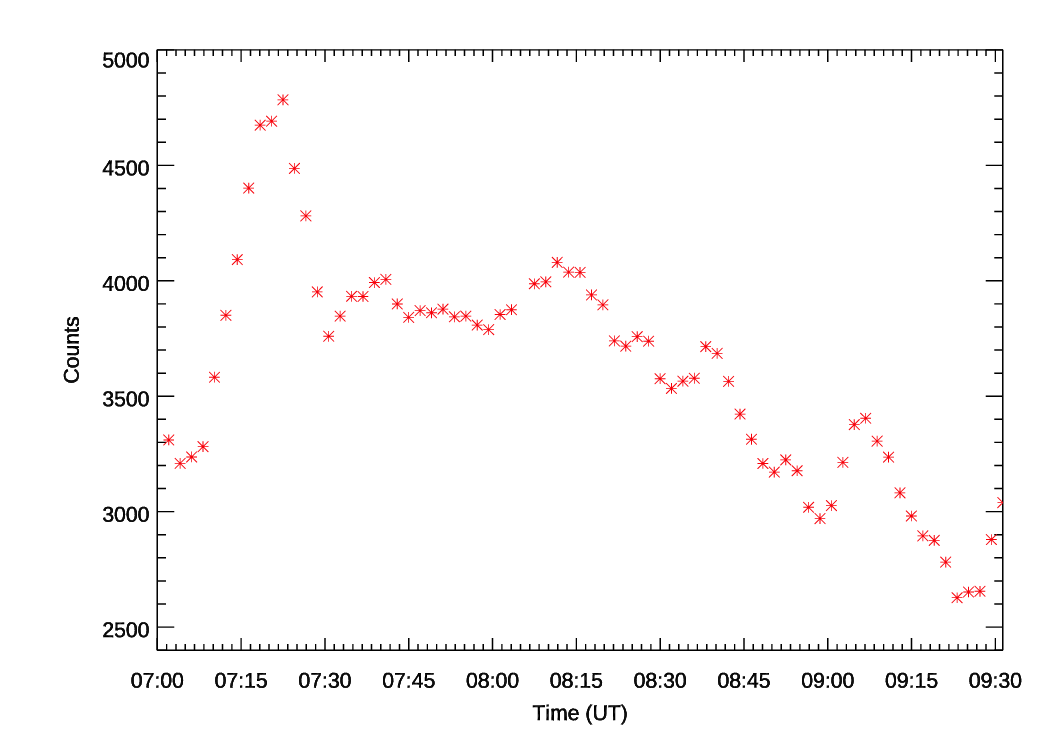}
\caption{Time variation of the peak intensity of the emission line, spatially averaged over the region of enhanced emission in Figure \ref{figone}. The position angle range is ${\approx}$\,98$^{\circ}$\,-\,101$^{\circ}$.}
\label{figtwo}
\end{figure}

Figure \ref{figone} shows the 5303\,{\AA} spectra constructed using the peak intensity of the emission line for the spectra due to the 1st slit.
The x-axis in the image corresponds to time 07:00\,-\,09:30\,UT. 
The y-axis is the solar position angle 
(${\theta}$), measured counter clockwise from north through east. It corresponds to the range $90^{\circ}$\,-\,$104^{\circ}$ for these observations.
Figure \ref{figtwo} shows the variation of the 
5303{\AA} emission line intensity in Figure \ref{figone} as a function of time, averaged over the position angle range ${\approx}\,98^{\circ}$\,-\,$101^{\circ}$.
Enhanced emission with onset at ${\approx}$07:04\,UT can be clearly seen. 
Though it decreased at ${\approx}$07:30\,UT, the background remained high till
${\approx}$08:45\,UT. 
Note that the average heliocentric distance of the coronal regions between position angles $98^{\circ}$ \& $101^{\circ}$ is $r\,{\approx}$\,1.18$R_{\odot}$ as compared to 
$r$\,=\,1.13$R_{\odot}$ for the center of the 1st slit of the spectrograph. This difference is due to the straight nature of the slit (see, e.g. Figure 2 in \citealp{Ramesh2024}). 

There was a X2.9 class GOES soft X-ray flare on 27 May 2024 from the heliographic location S18E89 during the period 06:49\,-\,07:25\,UT with peak at 07:08\,UT\footnote{https://www.solarmonitor.org/?date=20240527}. The flare was associated with a `halo' CME\footnote{https://cdaw.gsfc.nasa.gov/CME{\_}list/index.html}. 
The central position angle of the CME when it first appeared in the LASCO-C2 FoV at 
${\approx}$07:24:05\,UT is ${\approx}81^{\circ}$. The other observables of the CME during that time are angular width ${\approx}60^{\circ}$, and location of the leading edge 
at $r{\approx}4.07R_{\odot}$. CMEs with widths ${\leq}120^{\circ}$
are considered as non-halo events (see, e.g. \citealp{Webb2012}). The categorization of the above mentioned CME as a `halo' CME in the catalog is based on measurements at ${\approx}$08:36\,UT by which time it had expanded to be classified as a `halo' CME. Note that the width of a
CME is deﬁned as the maximum angle subtended by a CME on
the center of the Sun when the CME enters the LASCO-C3 FoV (see, e.g. \citealp{Pant2021}).
The position angle of the enhanced emission observed with the VELC is well within the angular extent of the CME.  It is well known that CMEs are primarily density enhancements above the background corona, similar to streamers \citep{Kathiravan2002,Ramesh2021}.  Hence, the changes in the line intensity in the present case are likely due to the CME \citep{Hori2005}, and mostly its leading edge \citep{Srivastava2000}. 

The duration of 
${\approx}$2\,h for which the observed intensity level remained higher compared to the background is in the range of time scales in which reconfiguration of the coronal magnetic field occurs in the aftermath of a CME \citep{Kathiravan2007}. If the enhancement observed with the VELC is due to the same CME observed in the LASCO-C2 FoV, then the speed of the CME between its above mentioned first appearances in the VELC and LASCO-C2 FoV should be 
${\approx}$1669\,$\rm km\,s^{-1}$. The extrapolation of the CME speeds at different epochs in the distance range $r{\approx}$4.07\,-\,27.90$R_{\odot}$ (using LASCO-C2 \& C3 observations) indicate a speed of 
${\approx}$1400\,$\rm km\,s^{-1}$ near the Sun. There is reasonable agreement between the two estimates considering that the impulsive acceleration phase of the CMEs occur close to the Sun, well below the lower limit of the LASCO-C2 FoV \citep{Zhang2001}. The maximum acceleration of a CME generally occurs at 
$r{\lesssim}1.4R_{\odot}$ \citep{Temmer2010}. Statistical studies indicate that the  acceleration phase of the CMEs end near the peak time of the associated soft X-ray flare \citep{Zhang2006}. These results suggest that the VELC and LASCO-C2 observations in the present case correspond most likely to the same CME. 

\subsection{Gauribidanur}

\begin{figure*}
\includegraphics[height=0.3\textheight]{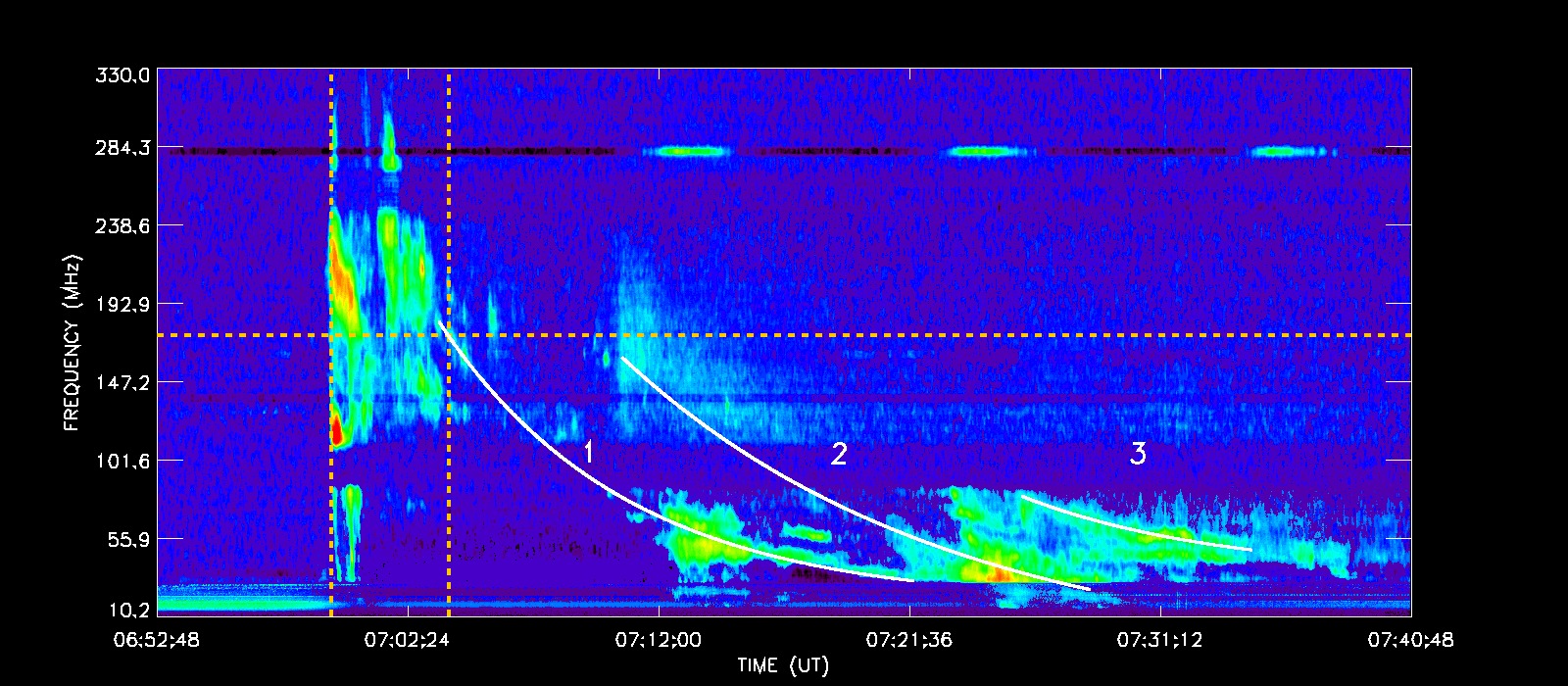}
\caption{Dynamic spectrum of type III and II bursts observed with GLOSS and GRASP on 2024 May 27 in the frequency range 330\,-\,30\,MHz. 
The horizontal dotted line is at 174\,MHz on the y-axis. The 1st and 2nd vertical dotted lines are at 06:59:30\,UT \& 07:04\,UT on the x-axis, respectively.
The fast drifting features during the interval 
${\approx}$06:59:30\,-\,07:03:30\,UT 
are type III bursts. The three comparitively slower drifting features marked 1, 2, \& 3 in the period  
${\approx}$07:04\,-\,07:40\,UT are the three type II bursts mentioned in the main text. The white lines shown are drawn by joining the mid-points in the respective slow drifting features, at each time interval.
The gap in the observed emission in the frequency range ${\approx}$108\,-\,88\,MHz is because of radio frequency interference (RFI) caused by FM transmission and the use of band rejection filter to suppress it. The horizontal patches near 138\,MHz \& 284\,MHz are also due to RFI. Otherwise the observing band is relatively free of RFI, which is one of the advantages with observations from Gauribidanur \citep{Monstein2007}.}
\label{figthree}
\end{figure*}

Radio spectral observations were carried out with Gauribidanur LOw-frequency Solar Spectrograph (GLOSS, \citealp{Kishore2014}) and Gauribidanur RAdio SpectroPolarimeter (GRASP, \citealp{Kishore2015,Hariharan2016b,Anshu2017b}) on 2024 May 27 around the same time as the VELC observations in the previous section. Figure \ref{figthree} shows the dynamic spectrum obtained with GLOSS and GRASP during the time 06:52:48\,-\,07:40:48\,UT on that day. Observations in the frequency range 330\,-\,30\,MHz and 30\,-\,10\,MHz are due to GLOSS and GRASP, respectively. In addition to fast drifting type III bursts, three slow drifting type II bursts 
can be seen in the frequency range ${\approx}$200\,-\,10\,MHz. 
The type III bursts seem to be in two groups. The 1st group is in the time period 
${\approx}$06:59:30\,-\,07:01:30\,UT. The start frequency in the beginning is close to ${\approx}$330\,MHz, but subsequently it decreases to
${\approx}$240\,MHz. The bursts are observed till 
${\lesssim}$30\,MHz on the low frequency side.  The 2nd group immediately follows the 1st group and lasts till 
${\approx}$07:03:30\,UT. The start frequency of the bursts decreases from 
${\approx}$330\,MHz to ${\approx}$240\,MHz in the case also. The low frequency limit in our observations is ${\approx}$70\,MHz.

\begin{figure}
\includegraphics[width=\columnwidth]{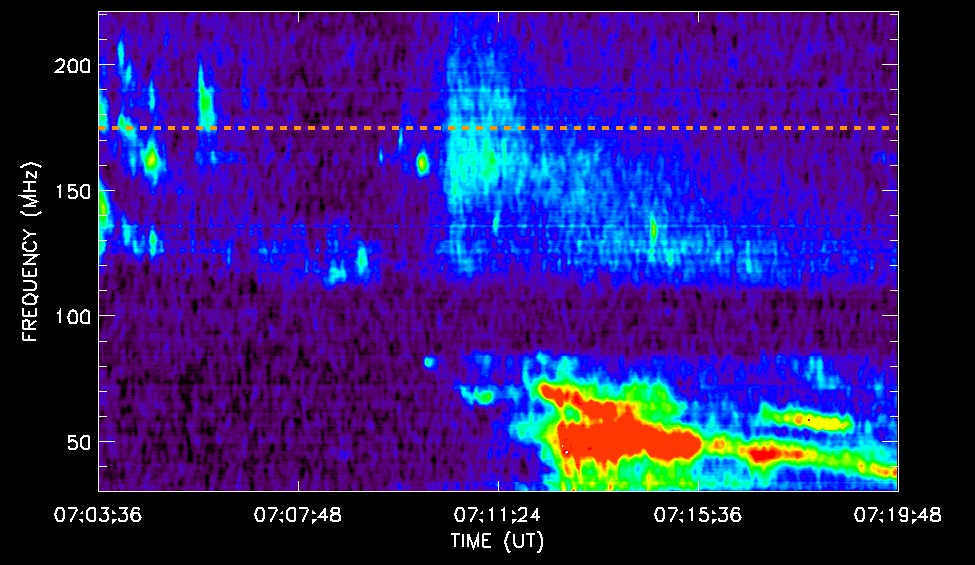}
\caption{Close-up view of Figure \ref{figthree}. Both the time and frequency ranges in the respective axes are restricted to see more clearly the patchy and weak emission in the frequency range 174\,-\,108\,MHz for the first type II burst mentioned in the main text (Section 2.2). The horizontal dotted line indicates 174\,MHz on the frequency axis.}
\label{figfour}
\end{figure}

A closer inspection of Figure \ref{figthree} indicates that the first 
type II 
burst is near
${\approx}$174\,MHz at 07:04\,UT (see Figure \ref{figfour}). 
The burst appears patchy in the frequency range 
174\,-\,108\,MHz. Furthermore, it shows one component near 
${\approx}$174\,MHz, and the other at ${\approx}$130\,MHz. These are not uncommon features in the dynamic spectra of type II bursts \citep{Carley2021}. 
We find similar discontinuous pattern in the dynamic spectra of the event obtained with Learmonth 
solar radio spectrograph\footnote{downloads.sws.bom.gov.au/wdc/wdc{\_}spec/data/learmonth/raw/24/} also. These correspondences suggest a common solar origin. In Figure \ref{figthree},
the frequency range of the
type II burst at ${\approx}$07:04\,UT is 
${\approx}$174\,-\,130\,MHz,
unlike the preceding 2nd group of type III bursts mentioned above. 
The instantaneous bandwidth of the type III bursts at ${\approx}$07:03\,UT
is ${\gtrsim}$170\,MHz. This is very large compared to the typical values for coronal type II radio bursts, ${\approx}$32\% of the center frequency  \citep{Mann1995,Ramesh2022}. 
Though patchy, a gradual drift towards lower frequencies could be also noticed in the close-up view of the type II burst in Figure \ref{figfour}. These chracteristics 
help to distinguish the type II burst in the present case.

The start times of the second and third type II bursts are ${\approx}$07:10\,UT and 07:26\,UT, respectively. 
Multiple type II bursts are considered to be caused by shock associated with the nose or leading edge of the CME (the fastest part in a CME), and interaction of the relatively slower CME flank with other coronal structures during its expansion \citep{Robinson1982,Cho2011}. 
Combining observations of a type II burst with three emission lanes and
extreme ultraviolet (EUV) images, 
\cite{Zimovets2015} showed that different lanes of the burst 
are due to propagation of different parts of 
the CME through coronal regions with different physical conditions. \cite{Alissandrakis2021} reported a similar type II burst with three bands, and related them with different parts of a propagating EUV wave.
Statistical studies 
indicate that the type II bursts can be located upto $46^{\circ}$ away from the leading edge of the CME \citep{Ramesh2012a}. The delayed occurrence of the second and third type II bursts in the present case are consistent with this scenario. There is a close association between the onset time of the first type II burst and that of the enhanced emission observed with VELC. So, we will discuss the first type II burst here. 

The location of the plasma level corresponding to 174\,MHz is 
$r{\approx}1.19R_{\odot}$ according to the two-fold electron density 
($N_{e}$) model of \cite{Newkirk1961}; also see \cite{Dulk1974}. We assumed the above density model since it is considered as a good representative of the active region corona associated with solar radio bursts \citep{Stewart1976,Vrsnak2004}. It has been used to successfully explain the drift rates and heights of type II bursts \citep{Mann1995,Klassen2002,Alissandrakis2021}.
The electron density corresponding to 174\,MHz is
$\rm {\approx}3.7{\times}10^{8}cm^{-3}$. This is typical of the density in the CMEs at the above distance  \citep{Akmal2001,Zucca2014_1,Anshu2019,Sheoran2023}. Observations in the 5303{\AA} emission line also indicate similar density \citep{Suzuki2006}.

Comparing the VELC and radio observations, we find that at ${\approx}$07:04\,UT the locations of the CME and first type II burst are ${\approx}$1.18$R_{\odot}$ and
${\approx}$1.19$R_{\odot}$, respectively. The associated active region is at the limb of the Sun (Section 2). So, any projection effects on the CME location are expected to be very minimal. The CME is likely to have propagated radially since the position angle of the associated enhanced emission in the VELC observations (Figure \ref{figone}) does not show any changes \citep{Muthupriyal2025}. 
The comparatively larger speed of the CME in the present case also indicates a radial propagation \citep{Gui2011}. The location of the CME source region at the limb (E89), and the smaller angular width (${\approx}60^{\circ}$) of the CME even at $r{\approx}4.07R_{\odot}$ too imply a radial movement in the distance range 
${<}1.2R_{\odot}$ mentioned above \citep{Burkepile2004}.   
The distances mentioned above are consistent with the general scenario that the shock is expected to be located ahead of the CME and
the type II burst is located at the shock front \citep{Gopalswamy2013}.
The separation between the CME leading edge and the shock, i.e. the shock standoff distance, can be  
${\approx}$0.01$R_{\odot}$ close to the Sun \citep{Gopalswamy2012}, same as in the present case.
This correlation indicates that the VELC CME, and first type II radio burst observed with the GLOSS and GRASP are closely associated. 
The empirical relation between CME height and type II burst 
frequency derived by \cite{Gopalswamy2013} indicate that for type II burst observations at 174\,MHz, the CME should have been at $r{\approx}1.16R_{\odot}$.
The close agreement beween this result, and the CME, type II burst locations ($1.18R_{\odot}$ and $1.19R_{\odot}$, respectively) mentioned above independently support the use of two-fold \cite{Newkirk1961} density model for the present work. 
We verified this using the present observations also.

The flux (F) of an optically thin spectral line at 
wavelength (${\lambda}$), for isothermal approximation, can be written as \citep{Aschwanden2005}, 
$F({\lambda})$\,=\,$G(T)N_{e}^{2}l$, where $G$ is the contribution function of the observed spectral line,
and $l$ is the depth along the line of sight.
The 5303\,{\AA} line is optically thin \citep{Mason1975}.   
The peak flux ($F$) corresponding to the CME associated enhanced emission in Figure \ref{figone} is ${\approx}10.3{\pm}0.9\, \rm erg\,sec^{-1} cm^{-2} sr^{-1}$.
This is in the range of reported values for the 5303\,{\AA} emission line flux \citep{Tsubaki1975}.
The value of $G$ for the
5303\,{\AA} solar coronal emission line, calculated using 
CHIANTI\footnote{http://www.solar.nrl.navy.mil/chianti.html}, 
is $1.87{\times}10^{-26}\, \rm erg\,sec^{-1} cm^{3} sr^{-1}$. 
The lateral width ($w$) of the enhancement in Figure \ref{figone} is 
${\approx}75^{\prime\prime}$.
Assuming CME plasma is isothermal \citep{Cheng2016,Sheoran2023,Debesh2025} and $l{=}w$, from the relationship mentioned above we find $N_{e}{\approx}2.2{\times}10^{8}\, \rm cm^{-3}$. The density corresponding to the plasma frequency of 174\,MHz mentioned earlier is reasonably close to the above value. This indicates that both the  type II radio burst and the enhanced 5303\,{\AA} emission in the present case are due to the same CME. 
Hence, the joint VELC-Gauribidanur observations reported in this work is an evidence for the occurrence of CME driven MHD shock close to the Sun at $r{\approx}1.19R_{\odot}$. \cite{Cliver2004} used SoHO/LASCO-C1 data to show that the 1997 November 6 CME was at  
$r{\approx}1.3R_{\odot}$ at the time of onset of the metric type II burst.
Statistical study using CME data obtained with the coronagraphs in the Mauna Loa Solar Observatory (MLSO) also indicate a similar result \citep{Cho2008}. Observations with COR1 coronagraph \citep{Howard2008}
onboard the Solar TErrestrial RElations Observatory A (STEREO-A)
indicate that the shocks occur at $r{<}1.5R_{\odot}$ \citep{Gopalswamy2009}. The above observations are in the visible wavelength range similar to VELC. Assuming that the EUV waves  
drive MHD shocks when associated with type II radio bursts, \cite{Gopalswamy2013} found that the shocks are located in the range $r{\approx}$1.20\,-\,1.93$R_{\odot}$.
The onset height of the type II burst reported in \cite{Zimovets2015} is comparitively lower ($r{\approx}$1.11$R_{\odot}$), but the location of the associated active region was 
${\approx}50^{\circ}$ inside the limb of the Sun. 

\section{Summary}

The observations reported indicate that by combining VELC data in the visible wavelength range and type II radio burst spectral observations it is possible to infer the shock formation very close to the Sun, chiefly due to the comparatively smaller size of the VELC occulter. In the present case, we find that the shock formed at a height of ${\approx}0.19R_{\odot}$ above the solar limb. 
The height of shock formation in the solar corona is important to understand particle acceleration by shocks, especially the release time of SEPs. 
The main requirement in this comparison is improved estimates of the CME location at the onset of the type II radio burst \citep{Mewaldt2012}.
The GLOSS and GRASP instruments observe the Sun for ${\approx}$10\,h everyday. 
Continuous observations of Sun with GLOSS, GRASP and similar radio spectrographs elsewhere are expected to facilitate comparision with VELC data, and that too 
at heights as small as $0.05R_{\odot}$ above the solar limb.
Joint observations with ADITYA-L1/VELC
and SDO/AIA in the future might help to compare the CME shock signatures in different wavelength ranges, particularly for events closer to the 
solar limb.

\section*{Acknowledgement}
The VELC team members who developed the payload are thanked for their efforts. ADITYA-L1 is an observatory class mission, funded and operated by the Indian Space Research Organization (ISRO). Data obtained with the different payloads onboard ADITYA-L1 are archived at the Indian Space Science Data Centre (ISSDC). We thank P. Savarimuthu, Priya Gavshinde, S. Nagashree, and E. Yuvashree for processing VELC data at the Payload Operations Center. We acknowledge the Gauribidanur Observatory team for
their help in the observations and upkeep of the facilities there. We thank the referees for insightful comments which helped to improve the manuscript.


\bibliography{reference} 


\end{document}